\documentclass[aps,prl,twocolumn,preprintnumbers,amsmath,amssymb,superscriptaddress]{revtex4-2}

\usepackage{amssymb}
\usepackage{mathtools}
\usepackage{physics}
\usepackage{graphicx}
\usepackage{hyperref}
\usepackage{slashed}
\usepackage{orcidlink}
\usepackage{bbm}
\usepackage{mathtools}
\usepackage{cancel}
\usepackage[caption=false]{subfig}

\usepackage{color}
\usepackage[export]{adjustbox}

\begin{document}

\author{Bilal Hawashin}
\email{hawashin@tp3.rub.de}
\affiliation{Theoretische Physik III, Ruhr-Universit\"at Bochum, D-44801 Bochum, Germany}

\author{Soumyadeep Chaudhuri}
\email{chaudhurisoumyadeep@gmail.com}
\affiliation{Physique Théorique et Mathématique,
Université Libre de Bruxelles,
C.P. 231, 1050 Brussels, Belgium}

\author{Igor F. Herbut}
\email{iherbut@sfu.ca}
\affiliation{Department of Physics, Simon Fraser University, Burnaby, British Columbia, Canada V5A 1S6}

\author{Michael M. Scherer}
\email{scherer@tp3.rub.de}
\affiliation{Theoretische Physik III, Ruhr-Universit\"at Bochum, D-44801 Bochum, Germany}

\title{All-temperature superfluidity in two dimensions}

\begin{abstract}
In local theories with short-range interactions, continuous symmetries cannot be spontaneously broken in two spatial dimensions at nonzero temperature. $\mathrm{U}(1)$-symmetric models instead admit a Berezinskii--Kosterlitz--Thouless (BKT) phase with quasi-long-range order and nonzero phase stiffness, which is normally destroyed at high temperatures by vortex proliferation. Here, we show that a BKT phase, and with it superfluidity, can persist to arbitrarily high temperatures. Using the large-$N$ expansion, we study a $(2+1)$-dimensional quantum field theory in which a complex scalar is coupled to the critical $\mathrm{O}(N)$ model. At the biconical fixed point, where the inter-sector coupling is negative, the thermal effective potential has a circle of minima with nonzero $\mathrm{U}(1)$ amplitude. Integrating out the gapped fluctuations yields an effective theory for the compact phase mode at distances large compared to $1/T$. Since temperature is the only scale at the fixed point, the phase stiffness grows linearly with temperature, with a proportionality factor that scales linearly with $N$ and is fixed entirely by the fixed point. This renders vortex operators irrelevant for sufficiently large $N$. We further show that the vortex fugacity is exponentially small in $N$.
\end{abstract}

\maketitle


\textit{Introduction. ---}
Spontaneous symmetry breaking is usually regarded as a low-temperature phenomenon. 
The standard argument is that at high temperatures the free energy is minimized by high-entropy configurations, which generally correspond to symmetry-preserving (disordered) phases. 
Recently, however, a series of works has established that spontaneous symmetry breaking can persist at arbitrarily high temperatures in certain ultraviolet-complete quantum field theories~\cite{Chai:2020symmetry,Chai:2020zgq,Buchel:2020thz,Chaudhuri:2020xxb,Chai:2020hnu,PhysRevD.103.096014,Chai:2021djc,Chaudhuri:2021dsq,Chai:2021tpt,Liendo:2022bmv,Ciccone:2023pdk,Hawashin:2025,KomargodskiPopov:2024,Bajc:2026infinite,Hawashin:2025ikp,Smolkin:2026wij,chaudhuri2026exploring,Smolkin:2026bxa,Harder:2026jjr}.
In many of these constructions, the phenomenon arises near the biconical fixed point~(BFP) of an $\mathrm{O}(N)\times\mathbb{Z}_2$-symmetric model, in which an $\mathrm{O}(N)$-symmetric sector with sufficiently large~$N$ is coupled to a $\mathbb{Z}_2$-odd scalar~$\chi$, and it is the discrete $\mathbb{Z}_2$ symmetry that remains broken at all temperatures~$T$~\cite{Chai:2020symmetry,Chai:2020zgq,Chai:2020hnu,Chai:2021djc,Hawashin:2025,Smolkin:2026wij}.
If the model is tuned such that the zero-temperature theory sits exactly at the BFP, then at any nonzero~$T$ the scalar~$\chi$ acquires a non-vanishing expectation value, which, since $T$ is the only scale, is fixed by dimensional analysis to be $\langle\chi\rangle_T \propto T^{\Delta_\chi}$, where~$\Delta_\chi$ is the scaling dimension of~$\chi$.

Spontaneous symmetry breaking at arbitrarily high temperatures has also been established in lattice models whose local degrees of freedom are unbounded, a mechanism termed \emph{entropic order}~\cite{entropicorder,Huang:2025gvi,Andriolo:2026udg,Hsin:2026ckp,Lucas:2026helical,Zane:2026entropic}. 
Similarly, for the ferromagnetic Ising model on planar lattices, arbitrarily high critical temperatures (relative to the exchange coupling) can be realized by making the maximal coordination number sufficiently large~\cite{Joseph:2026qxp}.
Models with short-range interactions, bounded local Hamiltonian terms, and continuous symmetries, however, cannot develop true long-range order at nonzero temperature in $d \leq 2$ spatial dimensions, as dictated by the Coleman--Hohenberg--Mermin--Wagner (CHMW) theorem~\cite{PhysRevLett.17.1133,PhysRev.158.383,Coleman:1973ci}.

\begin{figure}[t!]
    \centering
    \includegraphics[width=\columnwidth]{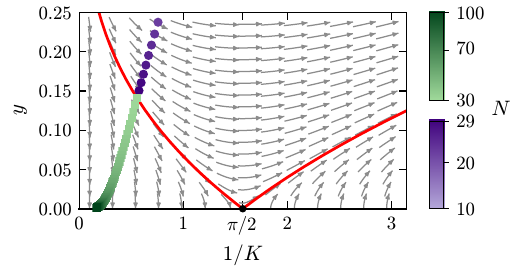}
    \caption{Kosterlitz--Thouless renormalization-group flow of $1/K$, with $K=\beta\rho_s$, and the vortex fugacity~$y$ of the effective action~\eqref{eq:effacSG}. Initial conditions below the separatrix (red) flow to the line of Gaussian fixed points ($y=0$, $1/K<\pi/2$), where vortices are irrelevant; those above flow to large~$y$, where vortices proliferate. The separatrix ends at the BKT transition, $1/K=\pi/2$. Symbols (colored by~$N$) show the matching value of~$K$ and the upper bound on~$y$ at $a\sim 1/T$ for the biconical fixed point at leading order in $1/N$. For sufficiently large~$N$, they lie deep in the region where the theory flows to a Gaussian theory with quasi-long-range order for $|x|\gg 1/T$.}
    \label{fig:BKTflow}
\end{figure}

Similarly, in local quantum field theories of complex scalars in two spatial dimensions, the global~$\mathrm{U}(1)$ symmetry cannot be spontaneously broken at any nonzero~$T$.
Nevertheless, it is well known that, while true long-range order is excluded, the theory of a single complex scalar exhibits quasi-long-range order at low $T$, characterized by an algebraic decay of correlation functions at long distances.
Upon increasing~$T$, the system undergoes the Berezinskii--Kosterlitz--Thouless (BKT) transition~\cite{Berezinskii:1970pzv,kosterlitz1973ordering} into a disordered phase with exponentially decaying correlations and a finite correlation length.
This transition is driven by the unbinding of vortex--antivortex pairs, which leads to a proliferation of free vortices.
The low-temperature BKT phase, in turn, is characterized by the irrelevance of vortex operators and a nonzero phase stiffness, which gives rise to superfluidity.

In this Letter, we show that a BKT phase can persist to arbitrarily high temperatures,
and that the underlying mechanism is the same fixed-point mechanism at work in the discrete case: $\mathrm{O}(N)\times\mathrm{U}(1)$ models possess a biconical fixed point at which, for sufficiently large $N$, the mixed coupling is negative.
At this fine-tuned point, fluctuations in the $\mathrm{O}(N)$ sector render the phase stiffness finite and large, and exponentially suppress the vortex fugacity, so that the unbinding of vortex--antivortex pairs becomes thermodynamically unfavorable at all temperatures.

Our main result is summarized in Fig.~\ref{fig:BKTflow}, which shows the Kosterlitz--Thouless renormalization-group flow in the plane spanned by $1/K$ and the vortex fugacity~$y$. Here, $K=\beta\rho_s$ is the dimensionless parameter, given by the product of the phase stiffness~$\rho_s$ and the inverse temperature~$\beta=1/T$.
Overlaid are our estimates of $K$ and the corresponding upper bounds on $y$ at the matching scale $a \sim 1/T$, obtained at the biconical fixed point. 
For sufficiently large $N$, $y$ is exponentially small, while the large value of $K$ renders the vortex perturbation strongly irrelevant. 
The theory therefore lies deep within the regime where it is described by a Gaussian theory for the phase mode at distances $|x| \gg 1/T$. 
We conclude that the thermal state of the underlying conformal field theory defined by the biconical fixed point exhibits quasi-long-range order at all temperatures.


\textit{Model and large-$N$ expansion. ---}
We consider a $(2+1)$-dimensional quantum field theory with global ${\mathrm{O}(N)\times\mathrm{U}(1)}$ symmetry, defined by the Euclidean action
\begin{align} \label{eq:actionmic}
    S =&\! \int d^3 x \bigg[\frac{1}{2} (\partial_\mu \phi_a)^2 + |\partial_\mu \psi|^2 + \frac{1}{2} m_\phi^2 \phi_a^2 + m_\psi^2 |\psi|^2 \nonumber \\
    & + \frac{\lambda_\phi}{8N} (\phi_a^2)^2 + \frac{\lambda_{\phi \psi}}{2N} \phi_a^2 |\psi|^2 + \frac{\lambda_\psi}{4N} |\psi|^4\bigg]+\dots\,,
\end{align}
where $\dots$ denote terms of higher order in the fields. Here, $\phi=(\phi_a)$, with $a=1,\dots,N$, is a real $\mathrm{O}(N)$ vector field, and $\psi$ is a complex scalar field charged under $\mathrm{U}(1)$.

Following Refs.~\cite{KomargodskiPopov:2024,chaudhuri2026exploring,Smolkin:2026wij,Smolkin:2026bxa}, we analyze the model at large but finite~$N$.
Our starting point is the $\mathrm{O}(N)$ vector model, defined by the action
\begin{align}
    S = \int d^3 x \bigg[ &\frac{1}{2} (\partial_\mu \phi_a)^2  + \frac{1}{2} m_\phi^2 \phi_a^2 + \frac{\lambda_\phi}{8N} (\phi_a^2)^2 \bigg].
\end{align}
A standard approach at large~$N$ is to introduce a Hubbard--Stratonovich field~$\sigma$~\cite{PhysRevD.10.2491,Moshe:2003xn,goykhman2020vector}, which brings the action into the form
\begin{align}
    S =\!\int d^3 x \bigg[ &\frac{1}{2} (\partial_\mu \phi_a)^2 + \frac{1}{2} \sigma \phi_a^2 + \frac{N m_\phi^2}{\lambda_\phi} \sigma - \frac{N}{2\lambda_\phi}  \sigma^2 \bigg],
\end{align}
where we additionally path integrate over $\sigma$ along the imaginary axis. 
Since $\phi_a$ now appears only quadratically in the action, we can integrate out the $N-1$ components $\phi_2,\dots,\phi_N$, which yields
\begin{align}
    S =& \frac{N-1}{2} \tr \log(-\Delta + \sigma) \\
    &+\int d^3 x \bigg[\frac{1}{2} (\partial_\mu \phi_1)^2 + \frac{1}{2} \sigma \phi_1^2 + \frac{N m_\phi^2}{\lambda_\phi} \sigma -   \frac{N}{2\lambda_\phi}  \sigma^2 \bigg].\notag
\end{align}
The first term in the above equation is ultraviolet divergent and must be regularized and renormalized; see, e.g., Ref.~\cite{Moshe:2003xn} for details. 
Apart from a field-independent constant, the only power-law divergence is linear in~$\sigma$ and can therefore be absorbed into~$m_\phi^2$. 
In the following, the trace-log term is understood to be renormalized in this way.
When tuned to criticality, the $(2+1)$-dimensional $\mathrm{O}(N)$ model flows to the interacting Wilson--Fisher fixed point, at which the large-$N$ expansion yields the leading-order scaling dimensions~\cite{PhysRevLett.29.1311,goykhman2020vector}
\begin{align}
    \Delta_\sigma &= 2 - \frac{32}{3 \pi^2 N},\qquad
    \Delta_\phi = \frac{1}{2} + \frac{4}{3 \pi^2 N}\,.
\end{align}
At large but finite $N$, the operator $\sigma^2$ is irrelevant. 
Hence, tuning only $m_\phi^2$ suffices to reach the fixed point, whose action is given by
\begin{align}
   \hspace{-0.15cm} S_N = \frac{N\!-\!1}{2} \tr \widetilde{\log}(-\Delta\! +\! \sigma)\! +\!\frac{1}{2}\! \int\! d^3 x \big[(\partial_\mu \phi_1)^2\! +
   \!\sigma \phi_1^2 \big],
\end{align}
where $\widetilde{\ldots}$ indicates that the field-independent term has been dropped and the linear term in an expansion in powers of $\sigma$ is removed by tuning to the critical point.

Here, we add a free complex field $\psi$ with scaling dimension $\Delta_\psi = 1/2$ and couple it to the critical $\mathrm{O}(N)$ sector through the operator $\sigma |\psi|^2$, which---at large but finite~$N$---is a slightly relevant deformation of the Wilson--Fisher fixed point coupled to a free scalar~\footnote{This is different to Ref.~\cite{KomargodskiPopov:2024}, where a $\mathbb{Z}_2$-odd scalar is added.}.
We then also have to consider the marginal self-interaction $|\psi|^6$.
The full perturbed $\mathrm{O}(N)\times\mathrm{U}(1)$-symmetric action is then
\begin{align}\label{eq:OnO2action}
    S= S_{N}+\int d^3 x \bigg[|\partial_\mu \psi|^2+ t \sigma |\psi|^2 + \frac{g}{6 N^2} |\psi|^6\bigg].
\end{align}


\textit{Beta functions. ---}
In the large-$N$ expansion, the beta functions of the theory in Eq.~\eqref{eq:OnO2action} read
\begin{align}
    \beta_t &= \frac{32}{3 \pi^2 N} t (t^2 - 1) + \mathcal{O}(1/N^2), \\
    \beta_g &= \frac{128}{\pi^2N} t^2 g - \frac{16384}{\pi^2 N} t^3 (t-1)^3 + \mathcal{O}(1/N^2)\,.
\end{align}
They follow from standard large-$N$ techniques~\cite{zinn2021quantum,Moshe:2003xn}.
More concretely, the beta functions can be directly obtained from the ones derived in our Ref.~\cite{chaudhuri2026exploring}, where $\mathrm{O}(N) \times \mathbb{Z}_2^M$-symmetric models have been considered within the large-$N$ expansion. 
The special case of $M = 2$, where we additionally restrict the couplings such that the symmetry is enhanced from $\mathbb{Z}_2^2$ to $\mathrm{O}(2) \cong \mathrm{U}(1)$, is precisely the model we consider here. See also~\footnote{At leading order in $1/N$, the diagrams contributing to $\beta_t$ and $\beta_g$ are identical to those of the model in Ref.~\cite{KomargodskiPopov:2024}; in particular, they are insensitive to the number of components of the scalar coupled to the $\mathrm{O}(N)$ sector, two for our $\mathrm{U}(1)$ field versus one for the $\mathbb{Z}_2$-odd field.} for an independent cross-check.

The scaling dimension of~$\psi$ reads~\cite{KomargodskiPopov:2024,chaudhuri2026exploring}
\begin{align}
    \Delta_\psi = \frac{1}{2} + \frac{4 t^2}{3 \pi^2 N} + \mathcal{O}(N^{-2}).
\end{align}
The above beta functions feature two interacting fixed points. 
In particular, apart from the decoupled solution $t_*=0$, $\beta_t = 0$ implies $t_* = \pm 1$. 
For $t_*=+1$, $\beta_g = 0$ implies $g_*=0$, and, after decomposing ${\psi = (\psi_1 + i \psi_2)/\sqrt{2}}$ with real $\psi_{1,2}$, it is evident that at this fixed point, the symmetry is enhanced to $\mathrm{O}(N+2)$. 
This is the isotropic fixed point~(IFP). For $t_* = -1$, we find $g_* = 1024$, which is the biconical fixed point. 
Both fixed points are well known in the context of multi-critical phenomena; see, e.g., Refs.~\cite{liu1973quantum,Fisher:1974,Kosterlitz:1976zz,PelissettoVicari:2002,Vicari:2003,Eichhorn:2013mcl}.


\textit{Saddle-point equations. ---}
Above, we have identified two fixed points with the fixed-point action
\begin{align}\label{eq:fpaction}
S_*= S_{N}+\int d^3 x \bigg[|\partial_\mu \psi|^2+ t_* \sigma |\psi|^2 + \frac{g_*}{6 N^2} |\psi|^6\bigg].
\end{align}
We now determine the ground state of the theory by minimizing the effective potential. At leading order in $1/N$, the effective potential is given by Eq.~\eqref{eq:fpaction} evaluated at constant field configurations, i.e., 
\begin{align}
    V_\mathrm{eff} = &\frac{N-1}{2 \Omega} \tr \widetilde{\log}(-\Delta + \sigma)\notag \\
    &+ \frac{1}{2} \sigma \phi_1^2 + t_* \sigma |\psi|^2 + \frac{g_*}{6 N^2} |\psi|^6,
\end{align}
where $\Omega$ is the spacetime volume. 
After defining $\phi_1 = N^{1/2} \bar{\phi}$, and $\psi = N^{1/2} \bar{\psi}$, the saddle-point equations read
\begin{align}
    \sigma \bar{\phi} &= 0, \label{eq:sad1}\\
    (-\Delta + \sigma)^{-1}_{x,x} + \bar{\phi}^2 + 2t_* |\bar{\psi} |^2 &= 0, \label{eq:sad2}\\
    (t_* \sigma + \frac{g_*}{2} |\bar{\psi}|^4) |\bar{\psi}| &= 0. \label{eq:sad3}
\end{align}
At zero temperature, the only physical ground state is $\phi = \psi = \sigma = 0$, as required by scale invariance. 
Explicitly, we can evaluate
\begin{align}
   (-\Delta + \sigma)_{x,x}^{-1} = - \frac{\sigma^{1/2}}{4 \pi}.
\end{align}
The first equation~\eqref{eq:sad1} implies $\sigma = 0$ or $\bar{\phi} = 0$. For $\sigma = 0$, Eq.~\eqref{eq:sad3} implies $\bar{\psi} = 0$, and together with Eq.~\eqref{eq:sad2} it follows $\bar{\phi} = 0$. Similarly, for $\bar{\phi} = 0$, Eqs.~\eqref{eq:sad2} and~\eqref{eq:sad3} imply $\sigma = \bar{\psi}  = 0$.

At nonzero temperature $T = 1/\beta$, the saddle-point equations~\eqref{eq:sad1} and~\eqref{eq:sad3} remain the same, but Eq.~\eqref{eq:sad2} is modified, since
\begin{align}
    \hspace{-0.15cm}(-\Delta + \sigma)_{x,x}^{-1}  = \frac{-1}{2\pi \beta} \log 
    \big[ 2\sinh\big( \frac{\beta \sqrt{\sigma}}{ 2} \big)\big] .
\end{align}
In contrast to the zero-temperature case, this term diverges when $\sigma \to 0$.
Hence, together with Eq.~\eqref{eq:sad1}, the only physical solution has $\bar{\phi} = 0$.
This is consistent with the CHMW theorem, according to which massless Goldstone fluctuations drive the thermal expectation value of~$\phi$ to zero at any nonzero temperature and for any $N > 1$ in $2+1$ dimensions.
For $\bar\phi = 0$, the remaining saddle-point equations can be solved numerically;
we find the symmetric saddle~($S_1$),  
\begin{align}
    \bar{\phi} = \bar{\psi} = 0, \quad \sigma \approx 0.926/\beta^2,
\end{align}
which exists for both the IFP and the BFP, while for the BFP with $t_* = -1$, we additionally find the solution~($S_2$),
\begin{align} \label{eq:sadsol}
    \bar{\phi} = 0, \quad |\bar{\psi}|^2 \approx 0.030/\beta, \quad \sigma \approx 0.455 /\beta^2.
\end{align}
Notice that the saddle-point solutions scale exactly with their respective large-$N$ scaling dimensions, as required by dimensional analysis. 
We emphasize that the existence of the saddle~$S_2$ relies on $t_* = -1$ and is hence a special feature of the BFP.

To determine which of the two saddles has the lower free energy at the BFP, we compare the effective potential at $S_1$ and $S_2$:
\begin{align}
    \frac{\Delta V_\mathrm{eff}}{N} &= \frac{1}{N} (V_\mathrm{eff}^{S_2} - V_\mathrm{eff}^{S_1}) \\
    &= \frac{1}{2\Omega} \tr \log\! \frac{-\Delta + \sigma_{S_2}}{-\Delta + \sigma_{S_1}} \!-\! \sigma_{S_2} \bar{|\psi|}^2_{S_2}\! +\! \frac{1024}{6} \bar{|\psi|}^6_{S_2}.\nonumber 
\end{align}
Since $\sigma_{S_2} < \sigma_{S_1}$, all eigenvalues of the operator in the argument of the logarithm are smaller than one, and hence the first term is negative.
By inserting the saddle-point values, we find that the sum of the last two terms is also negative.
We therefore conclude that $\beta^3 \Delta V_\mathrm{eff}/N < 0$, i.e., the saddle~$S_2$ has a lower free energy than~$S_1$.


\textit{Ground-state analysis at finite temperature. ---}
In contrast to the $\mathrm{O}(N)\times\mathbb{Z}_2$-symmetric model of Ref.~\cite{KomargodskiPopov:2024}, where the minima are two isolated points, the solution~\eqref{eq:sadsol} parametrizes a continuous circle of degenerate minima of the effective potential. 
Naively, each point on this circle would imply spontaneous breaking of the $\mathrm{U}(1)$ symmetry, which is however excluded by the CHMW theorem. 
In contrast, the leading-order saddle point is  manifestly compatible with the theorem in the $\mathrm{O}(N)$ sector because the $N-1$ Goldstone fluctuations contribute at $\mathcal{O}(N)$ and are thus already included.

We parametrize the complex field~$\psi$ by an amplitude mode~$\rho$ and a phase mode~$\theta$,
\begin{align}
    \psi(x) = \sqrt{\rho(x)} e^{i \theta(x)}\,.
\end{align}
Note that the field $\theta(x)$ appearing above is compact, i.e., $\theta \sim \theta + 2 \pi$. 
We then introduce the fluctuating fields
\begin{align}
     \rho(x) &= N \rho_0 + N^{1/2}\delta \rho(x),\qquad \phi(x) = \delta \phi(x),\\
    \sigma(x) &= \sigma_0 + N^{-1/2}\delta \sigma(x),
\end{align}
where $\rho_0$ and $\sigma_0$ solve the saddle-point equations.

The normalizations are chosen such that the two-point functions of $\delta\rho$, $\delta\phi$, and~$\delta\sigma$ are $\mathcal{O}(1)$. We (formally) integrate out all massive modes $\delta\rho$, $\delta\sigma$, and~$\delta\phi$, which yields an effective action for the phase mode~$\theta$ that is local on scales $|x|\gg 1/T$. In this regime, we employ a derivative expansion, whose leading term reads
\begin{align}\label{eq:effact}
    S_\mathrm{eff} = \frac{K}{2} \int d^2 x (\partial_\mu \theta)^2\,,
\end{align}
with the dimensionless parameter~$K = \beta \rho_s$ and phase stiffness~$\rho_s$. 
At next-to-leading order in $1/N$, this can be explicitly verified by integrating out Gaussian fluctuations around the saddle. 
At leading order in the large-$N$ expansion around the saddle~$S_2$, we find $\rho_s = 2\langle |\psi|^2 \rangle$, and therefore
\begin{equation}\label{eq:Kparam}
    K = \beta \rho_s \approx 0.060\,N + \mathcal{O}(1)\,.
\end{equation}
We expect the form of the effective action in Eq.~\eqref{eq:effact} for $|x| \gg 1/T$ to hold to all orders in the large-$N$ expansion.
Only the dimensionless stiffness~$K$ in Eq.~\eqref{eq:Kparam} receives corrections, which are small at large but finite~$N$.


\textit{Vortex operators. ---}
So far, we have ignored the compactness of~$\theta$, which allows for singular configurations in which $\theta$ winds by~$2\pi$ around a point. To account for these point defects, known as vortices, the action in Eq.~\eqref{eq:effact} must be replaced by~\cite{herbut2007modern,zinn2021quantum}
\begin{align}\label{eq:effacSG}
    \hspace{-0.2cm} S_\mathrm{eff} \!= \!\!\int\!\! d^2x \bigg[ \frac{K}{2} (\partial_\mu \theta_\mathrm{sw})^2\! +\! 
    \frac{1}{8 \pi^2 K}(\partial_\mu \tilde{\theta})^2\!-\!\frac{2y}{a^2}\cos \tilde{\theta}\bigg],
\end{align}
where $y$ is the vortex fugacity. It is defined at a matching scale $a \sim 1/T$ and must be determined from the full theory in the vortex sector. In the next section, we will derive an upper bound on~$y$.

The angle~$\theta_\mathrm{sw}$ describes the regular spin-wave part, whereas the dual field~$\tilde{\theta}$ carries the vortex contribution, with $\cos\tilde{\theta}$ interpreted as vortex operator~\cite{zinn2021quantum}.
The spin-wave part alone leads (a)~to the fulfillment of the CHMW theorem and (b)~to an algebraic decay of correlation functions; see, e.g., Refs.~\cite{herbut2007modern,zinn2021quantum}.
While (a) is stable against the vortex perturbation, (b)~is not.
For $K > 2/\pi$, the vortex operator is an irrelevant perturbation. 
Provided that the the fugacity is small at the matching scale, the theory remains gapless with quasi-long-range order, which is the defining characteristic of the BKT phase.
For $K < 2/\pi$, the vortex operator is relevant and drives the system into a gapped phase with a finite correlation length and an exponential decay of correlations.
A nonzero fugacity also renormalizes~$K$, as described by the Kosterlitz--Thouless flow equations,
\begin{align}
    \partial_\ell K^{-1} &= 4 \pi^3 y^2 + \mathcal{O}(y^4),\label{eq:KTeq1} \\
    \partial_\ell y &= (2 - \pi K)y + \mathcal{O}(y^3),\label{eq:KTeq2}
\end{align}
with $\ell = \log(b/a)$, where $b \geq a$ is the running length scale.

In our case, the dimensionless~$K$ at the matching scale~$a$ depends solely on~$N$ and is fixed entirely by the BFP.
At leading order in the large-$N$ expansion, we found $K \approx 0.060\,N + \mathcal{O}(1)$, and hence $K > 2/\pi$ is fulfilled if $N > 10$.
Values of~$N$ close to this threshold may lie outside the regime of validity of the leading-order large-$N$ analysis. 
Nevertheless, we can conclude that vortex operators are irrelevant perturbations of the Gaussian theory at large but finite~$N$.
Higher-vorticity vortices correspond to higher harmonics $\cos q\tilde{\theta}$ with scaling dimensions $\Delta_q = \pi K q^2 \geq \pi K$~\cite{herbut2007modern}, and are hence strongly irrelevant for $K > 2/\pi$.

The same conclusion follows from the original, more heuristic argument of Kosterlitz and Thouless~\cite{kosterlitz1973ordering}. 
Using Eqs.~\eqref{eq:effact} and~\eqref{eq:Kparam}, we compare the energy of an isolated vortex with its entropy gain in a system of radial size~$R$,
\begin{align}
    \Delta F &\approx (\pi\rho_s - 2T) \log(R/a)\notag\\
    &\approx T(0.060\,\pi N - 2) \log(R/a),
\end{align}
up to an $R$-independent core contribution. Here, $a \sim 1/T$ is the short-distance cutoff of the effective description~\eqref{eq:effact}. 
Isolated vortices are thus unfavorable for $N > 10$, and, since $\Delta F \propto T$ at the fixed point, this holds at all temperatures.


\textit{Vortex fugacity and BKT phase. ---}
To establish that the thermal state exhibits a BKT phase at all temperatures, it remains to show that the vortex fugacity is small at large~$N$.
The fugacity is $y = y_0 \exp(-S_\mathrm{core})$, where $S_\mathrm{core}$ is the action of the core of a single unit vortex. 
It is evident that such a term can not be obtained from a large-$N$ expansion around the homogeneous saddle~\footnote{Equivalently, homogeneous configurations and vortices have different winding numbers and thus lie in different homotopy classes.}, so the vortex sector must be treated explicitly.
From inhomogeneous saddle-point solutions, we derive a lower bound on the vortex core action that scales linearly with~$N$.
At leading order in $1/N$, this bounds the fugacity by $y \leq A\,e^{-cN}$ with positive constants~$A$ and~$c$. 
For sufficiently large~$N$, $y$ is thus exponentially small, the model flows to the Gaussian theory for $|x|\gg 1/T$, and the thermal state exhibits quasi-long-range order at all temperatures.

We return to the fixed-point action~\eqref{eq:fpaction} and rescale $\psi = N^{1/2}\bar{\psi}$. At large~$N$, the path integral is dominated by saddle points.
Relaxing the restriction to homogeneous saddles, we allow arbitrary $\sigma(x)$ and $\psi(x)$, while setting $\phi=0$ as required by the CHMW theorem. 
The saddle-point equation for~$\sigma(x)$ then reads
\begin{align}\label{eq:sadsigma}
   (-\Delta + \sigma)^{-1}_{x,x} + 2t_* |\bar{\psi}|^2 = 0,
\end{align}
where the coincident propagator is now a nonlocal functional of~$\sigma(x)$. 
This makes a general solution difficult. However, using
\begin{align}
    \frac{\delta^2 S}{\delta \sigma(x) \delta \sigma(y)}
    &= -\frac{N}{2} G_\phi(x,y) G_\phi(y,x)
\end{align}
with $G_\phi(x,y) = (-\Delta + \sigma)^{-1}_{x,y}$, we can establish that, for fixed $\psi$, the functional $S[\psi, \sigma]$ is concave in $\sigma$.

To that end, consider the variation $\sigma_\epsilon(x) = \sigma(x) + \epsilon\, h(x)$ with real $h$. Then
\begin{align}
    &\frac{d^2}{d\epsilon^2} S[\psi, \sigma + \epsilon h]\bigg|_{\epsilon = 0} \!= \int d^3x d^3y\, \frac{\delta^2 S}{\delta \sigma(x) \delta \sigma(y)} h(x) h(y)\notag \\[5pt]
    &= -\frac{N}{2}\!\! \int d^3x d^3y \left(\sqrt{|h(x)|} G_\phi(x,y) \sqrt{|h(y)|}\right)^2\leq 0.
\end{align}
For any fixed field configuration~$\psi$, the action~$S$ is therefore concave in~$\sigma$, and hence $S[\psi,\tilde{\sigma}] \geq S[\psi, \sigma]$, where $\tilde{\sigma}(x)$ is any solution of Eq.~\eqref{eq:sadsigma}.
This, however, also implies that $S[\psi,\tilde{\sigma}] \geq S[\psi, \sigma_0]$, where $\sigma_0$ is the solution of the homogeneous saddle-point equation, cf. Eq.~\eqref{eq:sadsol}. We can use this property to compare the action evaluated at any saddle-point solution with the action at the homogeneous solution, i.e.,
\begin{align}
    \Delta S = S[\psi,\tilde{\sigma}] - S[\psi_0 ,\sigma_0] &\geq S[\psi, \sigma_0] - S[\psi_0 ,\sigma_0].
\end{align}
For the right-hand side, the log-term cancels, and we find
\begin{align}\label{eq:actionbound}
    \frac{\Delta S}{N} &\!\geq\! \int\! d^3x \bigg[|\partial_\mu \bar{\psi}|^2 \!+\! \frac{g_*}{6} (|\bar{\psi}|^2 \!-\! |\bar{\psi}_0|^2)^2(|\bar{\psi}|^2 \!+\! 2 |\bar{\psi}_0|^2)\bigg] \nonumber\\
    &\geq 0,
\end{align}
where $\bar{\psi}=\psi/\sqrt{N}$, $\ \bar{\psi}_0=\psi_0/\sqrt{N}$, and we have used that at the homogeneous saddle, $\sigma_0 = g_*|\bar{\psi}_0|^4/2$.
Any inhomogeneous saddle therefore has a higher free energy than the homogeneous one, and its contribution is suppressed at large~$N$. 
This also includes vortex configurations.


\textit{Unit-vortex strings. ---}
To estimate the vortex core energy explicitly, we follow Ref.~\cite{eto2009non} and take the following ansatz for a unit vortex, which for simplicity we assume to be independent of imaginary time (a ``vortex string''):
\begin{align}
    \psi(r,\varphi) = \sqrt{N\rho_0} f(r) e^{i\varphi}, \, f(0) = 0, \, \lim_{r\to \infty} f(r) = 1,
    \label{ansatz_nonhom}
\end{align}
where $N\rho_0 = |\psi_0|^2$. 
This configuration has winding number $q = (2\pi)^{-1}\oint_C d\vec{s}\cdot\vec{\nabla}\arg\psi = 1$, where $C$ is any counterclockwise loop encircling the origin, and thus lies in a different homotopy class than the homogeneous configuration with $q=0$. The boundary conditions ensure regularity at the core and that $|\psi|$ approaches its homogeneous value far from it.

We will continue by deriving a lower bound for the core energy of the unit-vortex configuration in Eq.~\eqref{ansatz_nonhom}. Plugging the ansatz in Eq.~\eqref{eq:actionbound} yields
\begin{align}
    \Delta S \geq 2\pi \beta \rho_0 N \!\int_0^\infty\! dr r\bigg[& f'^2 + \frac{f^2}{r^2} \nonumber\\
    &+ \frac{g_* \rho_0^2}{6} (f^2 \!-\! 1)^2(f^2\!+\!2)\bigg].
\end{align}
To bound this functional from below, we define ${r_0^2 = 6/(g_*\rho_0^2) \sim 1/T^2}$ and use $f'^2 \geq 0$, $f^2+2 \geq 2$, and $(f^2-1)^2 \geq (f-1)^2$, which yields
\begin{align}
    \Delta S \geq 2\pi \beta \rho_0 N \int_0^\infty dr r\left( \frac{f^2}{r^2} + \frac{2}{r_0^2} (f - 1)^2\right)\,.
\end{align}
Its stationarity condition, $f/r^2 + 2(f-1)/r_0^2 = 0$, has the solution $f(r) = 2r^2/(2r^2 + r_0^2)$, which is a minimizer and satisfies the boundary conditions imposed above.
For this configuration, we find
\begin{equation}
    \Delta S \geq \pi\beta\rho_0 N \int_0^{\infty} dr^2\, \frac{2}{2r^2 + r_0^2}.
\end{equation}
This integral diverges logarithmically, as is typical for isolated vortices in infinite systems. We therefore evaluate it up to a cutoff~$R$,
\begin{align}
    \Delta S &\geq \pi \beta \rho_0 N \!\int_0^{R^2}\! dr^2\! \frac{2}{2r^2\! +\! r_0^2} 
    \notag \\
    &= 2\pi \beta \rho_0 N \log(R/r_0) + \pi \beta \rho_0 N \log(2 + \frac{r_0^2}{R^2})\notag \\
    &\geq 2\pi \beta \rho_0 N \log(R/r_0) + \pi \beta \rho_0 N \log(2).
    \label{eq:estimatefug}
\end{align}
The last term is a lower bound on the vortex core action. 
To see this, note that outside the core, the effective action~\eqref{eq:effact} gives the standard long-range tail of a vortex,
\begin{align}
    S_\mathrm{vortex} = S_\mathrm{core}(a) + \pi \beta \rho_s \log(R/a),
\end{align}
where $a \sim 1/T$ is the short-distance cutoff introduced above.
The core action is thus defined by evaluating the vortex action on a disk of radius~$R$ and subtracting the long-range tail, i.e.,
\begin{align}
    S_\mathrm{core}(a) = S_\mathrm{vortex} - \pi \beta \rho_s \log(R/a),
\end{align}
which remains finite as $R\to\infty$. The result depends on the choice of~$a$: changing the cutoff to another $a' \sim 1/T$ gives $S_\mathrm{core}(a') - S_\mathrm{core}(a) = \pi \beta \rho_s \log(a'/a)$~\footnote{The fugacities at the matching scales~$a$ and~$a'$ lie on the same renormalization-group trajectory, cf. Eqs.~\eqref{eq:KTeq1} and~\eqref{eq:KTeq2}, and the factor of~2 from the dimensional rescaling can be absorbed into the prefactor~$y_0$ of $y=y_0\exp(-S_{\mathrm{core}})$.}.
Specifying the core action therefore requires specifying the cutoff. 
Choosing $a = r_0 \sim 1/T$, 
Eq.~\eqref{eq:estimatefug} yields
\begin{align}
    S_\mathrm{core}(r_0) \geq \pi \beta \rho_0 N \log(2) = \frac{\pi}{2} K \log(2) \approx 0.065 N,
    \label{eq:core action bound}
\end{align}
with $K$ as in Eq.~\eqref{eq:Kparam}.
Hence, at large but finite~$N$, the vortex fugacity $y \sim \exp(-S_\mathrm{core})$ is exponentially small in~$N$.

\begin{figure}[t!]
    \centering   \includegraphics[width=\columnwidth]{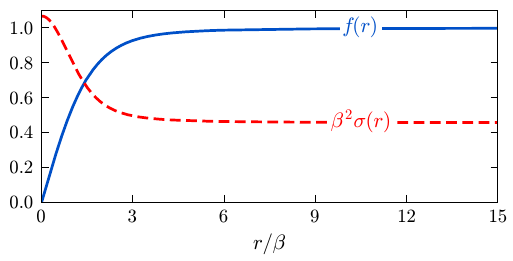}
    \caption{Numerical solutions of $f(r)$ and $\sigma(r)$ for the unit-vortex saddle in Eq.~\eqref{ansatz_nonhom}. 
    For large~$r$, it is evident that $\bar{\psi}$ and $\sigma$ approach their respective homogeneous saddle values in Eq.~\eqref{eq:sadsol}.}
 \label{fig:numerical_sols}
\end{figure}

We can additionally verify this by numerically solving the saddle-point equations, which consist of Eq.~\eqref{eq:sadsigma} together with
\begin{align}\label{eq:secondeqsadnonhom}
     -\Delta\bar{\psi}+t_*\sigma\bar{\psi}+\frac{g_*}{2}|\bar{\psi}|^4\bar{\psi}= 0,
\end{align}
using the unit-vortex ansatz~\eqref{ansatz_nonhom} for~$\psi$ together with a imaginary-time independent and radially-symmetric profile~$\sigma(r)$.
With this ansatz, we find a numerical solution for $f(r)$ and $\sigma(r)$, shown in Fig.~\ref{fig:numerical_sols}, which explicitly confirms the existence of unit-vortex saddles.
Plugging these solutions into the action and comparing it with the action at the homogeneous saddle, we find 
\begin{align}
    S_\mathrm{vortex} \approx 0.085 N + 0.187 N \log(R/\beta)
\end{align}
for large values of $R/\beta$. As before, we extract the core-contribution by subtracting the long-range tail, yielding $S_\mathrm{core}(\beta) \approx 0.085 N$.
Using the relation between the scales $r_0$ and $\beta$ as well as the relation between the values of $S_\mathrm{core}$ at these two scales, we get
 \begin{align}
    S_\mathrm{core}(r_0) \approx 0.085 N+0.187 N\log(r_0/\beta)\approx 0.260 N,
\end{align}
which also satisfies the bound in Eq.~\eqref{eq:core action bound}. This corroborates the exponential suppression of the vortex fugacity.

\textit{Conclusion. ---} We have shown that a BKT phase, and with it two-dimensional superfluidity, can persist at all temperatures in a local, unitary quantum field theory.
For the $\mathrm{O}(N)\times\mathrm{U}(1)$ model tuned to its biconical fixed point at zero temperature, the temperature is the only scale of the thermal state, so the dimensionless stiffness~$K$ and the vortex fugacity~$y$ at the matching scale $a\sim1/T$ are temperature-independent.
At large but finite~$N$, $K\propto N$ and $y$ is bounded by an exponentially small value, placing the theory deep within the BKT phase.

\textit{Acknowledgments. ---}  
MMS was funded by Deutsche Forschungsgemeinschaft (DFG) within Project-ID 277146847, SFB 1238 (project C02), and the DFG Heisenberg programme (Project-ID 452976698). IFH was supported by the NSERC of Canada. 
The work of SC is partially supported by the International Solvay Institutes and the Belgian Fonds National de la Recherche Scientifique FNRS (convention IISN 4.4503.15). 
SC is also supported by a postdoctoral research fellowship of FNRS. The numerical solution for the unit-vortex saddle was obtained with the help of OpenAI GPT-6 Astra.

\bibliographystyle{apsrev4-1}
\bibliography{references}

@article{Kosterlitz:1976zz,
    author = "Kosterlitz, J. M. and Nelson, David R. and Fisher, Michael E.",
    title = "{Bicritical and tetracritical points in anisotropic antiferromagnetic systems}",
    doi = "10.1103/PhysRevB.13.412",
    journal = "Phys. Rev. B",
    volume = "13",
    pages = "412--432",
    year = "1976"
}

@article{PelissettoVicari:2002,
    author = "Pelissetto, Andrea and Vicari, Ettore",
    title = "{Critical phenomena and renormalization group theory}",
    eprint = "cond-mat/0012164",
    archivePrefix = "arXiv",
    doi = "10.1016/S0370-1573(02)00219-3",
    journal = "Phys. Rept.",
    volume = "368",
    pages = "549--727",
    year = "2002"
}

@article{Harder:2026jjr,
    author = "Harder, Sven and Sharon, Adar",
    title = "{A Superfluid at All Temperatures}",
    eprint = "2610.10169",
    archivePrefix = "arXiv",
    primaryClass = "hep-th",
    month = "10",
    year = "2026"
}

@article{Zane:2026entropic,
    author = "Zane, Isabella and Lucas, Andrew",
    title = "{Entropic-ordered superfluid}",
    eprint = "2610.12635",
    archivePrefix = "arXiv",
    primaryClass = "cond-mat.stat-mech",
    month = "10",
    year = "2026"
}

@article{Lucas:2026helical,
    author = "Lucas, Andrew",
    title = "{Spontaneous symmetry breaking at any temperature in a local one-dimensional model}",
    eprint = "2610.07693",
    archivePrefix = "arXiv",
    primaryClass = "math-ph",
    month = "10",
    year = "2026"
}

@article{Joseph:2026qxp,
    author = "Joseph, Davidson Noby and Walsh, Connor M. and Boettcher, Igor",
    title = "{Families of planar lattices with arbitrarily high Tc for the ferromagnetic Ising model}",
    doi = "10.1103/y4zd-dffy",
    journal = "Phys. Rev. E",
    volume = "114",
    number = "2",
    pages = "024133",
    year = "2026"
}

@article{Moshe:2003xn,
    author = "Moshe, Moshe and Zinn-Justin, Jean",
    title = "{Quantum field theory in the large N limit: A Review}",
    eprint = "hep-th/0306133",
    archivePrefix = "arXiv",
    doi = "10.1016/S0370-1573(03)00263-1",
    journal = "Phys. Rept.",
    volume = "385",
    pages = "69--228",
    year = "2003"
}

@article{Bajc:2026infinite,
    author = "Bajc, Borut and Muco, Giulia and Sannino, Francesco and Wagner, Sophie",
    title = "{Infinite Heat Order in 3+1 Dimensions}",
    eprint = "2604.01184",
    archivePrefix = "arXiv",
    primaryClass = "hep-th",
    month = "4",
    year = "2026"
}

@article{Buchel:2020thz,
    author = "Buchel, Alex",
    title = "{Thermal order in holographic CFTs and no-hair theorem violation in black branes}",
    eprint = "2005.07833",
    archivePrefix = "arXiv",
    primaryClass = "hep-th",
    doi = "10.1016/j.nuclphysb.2021.115425",
    journal = "Nucl. Phys. B",
    volume = "967",
    pages = "115425",
    year = "2021"
}

@article{Smolkin:2026wij,
    author = "Smolkin, Michael and Yung, Lev",
    title = "{Thermal Order in the Biconical Model}",
    eprint = "2608.02720",
    archivePrefix = "arXiv",
    primaryClass = "hep-th",
    month = "8",
    year = "2026"
}

@article{Smolkin:2026bxa,
    author = "Smolkin, Michael and Yung, Lev",
    title = "{Persistent Spontaneous Time-Reversal Breaking}",
    eprint = "2609.16162",
    archivePrefix = "arXiv",
    primaryClass = "hep-th",
    month = "9",
    year = "2026"
}

@article{eto2009non,
  title="{Non-Abelian global vortices}",
  author="{Eto, Minoru and Nakano, Eiji and Nitta, Muneto}",
  journal="{Nuclear Physics B}",
  volume={821},
  number={1-2},
  pages={129--150},
  year={2009},
  publisher={Elsevier},
  doi={10.1016/j.nuclphysb.2009.06.013}
}

@article{PhysRevLett.29.1311,
  title = "{Critical Exponents for Charged and Neutral Bose Gases above $\ensuremath{\lambda}$ Points}",
  author = {Ma, Shang-keng},
  journal = {Phys. Rev. Lett.},
  volume = {29},
  issue = {19},
  pages = {1311--1314},
  numpages = {0},
  year = {1972},
  month = {Nov},
  publisher = {American Physical Society},
  doi = {10.1103/PhysRevLett.29.1311},
  url = {https://link.aps.org/doi/10.1103/PhysRevLett.29.1311}
}

@article{Hsin:2026ckp,
    author = "Hsin, Po-Shen and Kobayashi, Ryohei",
    title = "{Exploring Entropic Orders: High Temperature Continuous Symmetry Breaking, Chiral Topological States and Local Commuting Projector Models}",
    eprint = "2604.18694",
    archivePrefix = "arXiv",
    primaryClass = "cond-mat.str-el",
    month = "4",
    year = "2026"
}

@article{Andriolo:2026udg,
    author = "Andriolo, Enrico and Nguyen, Mendel and Richards, Emily and Sulejmanpasic, Tin",
    title = "{Proof of entropic order in Generalized Ising Models}",
    eprint = "2604.09768",
    archivePrefix = "arXiv",
    primaryClass = "cond-mat.stat-mech",
    month = "4",
    year = "2026"
}

@article{Huang:2025gvi,
    author = "Huang, Xiaoyang and Komargodski, Zohar and Lucas, Andrew and Popov, Fedor K. and Sulejmanpasic, Tin",
    title = "{Minimal Models of Entropic Order}",
    eprint = "2512.07980",
    archivePrefix = "arXiv",
    primaryClass = "cond-mat.stat-mech",
    month = "12",
    year = "2025"
}

@article{PhysRevD.10.2491,
  title = {Spontaneous symmetry breaking in the $\mathrm{O}(N)$ model for large $N$},
  author = {Coleman, Sidney and Jackiw, R. and Politzer, H. D.},
  journal = {Phys. Rev. D},
  volume = {10},
  issue = {8},
  pages = {2491--2499},
  numpages = {0},
  year = {1974},
  month = {Oct},
  publisher = {American Physical Society},
  doi = {10.1103/PhysRevD.10.2491},
  url = {https://link.aps.org/doi/10.1103/PhysRevD.10.2491}
}

@article{kosterlitz1973ordering,
  title={Ordering, metastability and phase transitions in two-dimensional systems},
  author={Kosterlitz, John Michael and Thouless, David James},
  journal={Journal of Physics C: Solid State Physics},
  volume={6},
  number={7},
  pages={1181--1203},
  year={1973},
  doi={10.1088/0022-3719/6/7/010}
}

@article{Berezinskii:1970pzv,
    author = "Berezinskii, V. L.",
    title = "{Destruction of Long-range Order in One-dimensional and Two-dimensional Systems having a Continuous Symmetry Group I. Classical Systems}",
    journal = "Sov. Phys. JETP",
    volume = "32",
    pages = "493--500",
    year = "1971",
  url = {http://www.jetp.ras.ru/cgi-bin/e/index/e/32/3/p493?a=list}
}

@book{zinn2021quantum,
  title={Quantum field theory and critical phenomena},
  author={Zinn-Justin, Jean},
  volume={171},
  year={2021},
  publisher={Oxford university press},
  doi={10.1093/oso/9780198834625.001.0001}
}

@article{goykhman2020vector,
  title={Vector model in various dimensions},
  author={Goykhman, Mikhail and Smolkin, Michael},
  journal={Physical Review D},
  volume={102},
  number={2},
  pages={025003},
  year={2020},
  publisher={APS},
  doi={10.1103/PhysRevD.102.025003}
}

@article{chaudhuri2026exploring,
  title={Exploring thermal order in conformal theories with multiple scalars coupled to an {$O(N)$} vector field},
  author={Chaudhuri, Soumyadeep and Hawashin, Bilal and Rabinovici, Eliezer and Scherer, Michael M.},
  eprint={2609.00160},
  archiveprefix={arXiv},
  primaryclass={hep-th},
  reportnumber={CERN-TH-2026-205},
  year={2026}
}

@article{PhysRevLett.17.1133,
  title = {Absence of Ferromagnetism or Antiferromagnetism in One- or Two-Dimensional Isotropic Heisenberg Models},
  author = {Mermin, N. D. and Wagner, H.},
  journal = {Phys. Rev. Lett.},
  volume = {17},
  issue = {22},
  pages = {1133--1136},
  numpages = {0},
  year = {1966},
  month = {Nov},
  publisher = {American Physical Society},
  doi = {10.1103/PhysRevLett.17.1133},
  url = {https://link.aps.org/doi/10.1103/PhysRevLett.17.1133}
}

@article{Coleman:1973ci,
    author = "Coleman, Sidney R.",
    title = "{There are no Goldstone bosons in two-dimensions}",
    doi = "10.1007/BF01646487",
    journal = "Commun. Math. Phys.",
    volume = "31",
    pages = "259--264",
    year = "1973"
}

@article{Chaudhuri:2020xxb,
    author = "Chaudhuri, Soumyadeep and Choi, Changha and Rabinovici, Eliezer",
    title = "{Thermal order in large N conformal gauge theories}",
    eprint = "2011.13981",
    archivePrefix = "arXiv",
    primaryClass = "hep-th",
    doi = "10.1007/JHEP04(2021)203",
    journal = "JHEP",
    volume = "04",
    pages = "203",
    year = "2021"
}

@article{PhysRevD.103.096014,
  title = {Asymptotically free and safe fate of symmetry nonrestoration},
  author = {Bajc, Borut and Lugo, Adri\'an and Sannino, Francesco},
  journal = {Phys. Rev. D},
  volume = {103},
  issue = {9},
  pages = {096014},
  numpages = {20},
  year = {2021},
  month = {May},
  publisher = {American Physical Society},
  doi = {10.1103/PhysRevD.103.096014},
  url = {https://link.aps.org/doi/10.1103/PhysRevD.103.096014}
}

@article{Chaudhuri:2021dsq,
    author = "Chaudhuri, Soumyadeep and Rabinovici, Eliezer",
    title = "{Symmetry breaking at high temperatures in large N gauge theories}",
    eprint = "2106.11323",
    archivePrefix = "arXiv",
    primaryClass = "hep-th",
    doi = "10.1007/JHEP08(2021)148",
    journal = "JHEP",
    volume = "08",
    pages = "148",
    year = "2021"
}

@article{PhysRev.158.383,
  title = {Existence of Long-Range Order in One and Two Dimensions},
  author = {Hohenberg, P. C.},
  journal = {Phys. Rev.},
  volume = {158},
  issue = {2},
  pages = {383--386},
  numpages = {0},
  year = {1967},
  month = {Jun},
  publisher = {American Physical Society},
  doi = {10.1103/PhysRev.158.383},
  url = {https://link.aps.org/doi/10.1103/PhysRev.158.383}
}

@article{Liendo:2022bmv,
    author = "Liendo, Pedro and Rong, Junchen and Zhang, Haoyu",
    title = "{Spontaneous breaking of finite group symmetries at all temperatures}",
    eprint = "2205.13964",
    archivePrefix = "arXiv",
    primaryClass = "hep-th",
    doi = "10.21468/SciPostPhys.14.6.168",
    journal = "SciPost Phys.",
    volume = "14",
    number = "6",
    pages = "168",
    year = "2023"
}

@article{Ciccone:2023pdk,
    author = "Ciccone, Riccardo and Di Pietro, Lorenzo and Serone, Marco",
    title = "{Anomalies and persistent order in the chiral Gross-Neveu model}",
    eprint = "2312.13756",
    archivePrefix = "arXiv",
    primaryClass = "hep-th",
    doi = "10.1007/JHEP02(2024)211",
    journal = "JHEP",
    volume = "02",
    pages = "211",
    year = "2024"
}

@article{Chai:2020zgq,
    author = "Chai, Noam and Chaudhuri, Soumyadeep and Choi, Changha and Komargodski, Zohar and Rabinovici, Eliezer and Smolkin, Michael",
    title = "{Thermal Order in Conformal Theories}",
    eprint = "2005.03676",
    archivePrefix = "arXiv",
    primaryClass = "hep-th",
    doi = "10.1103/PhysRevD.102.065014",
    journal = "Phys. Rev. D",
    volume = "102",
    number = "6",
    pages = "065014",
    year = "2020"
}

@article{Chai:2021djc,
    author = "Chai, Noam and Dymarsky, Anatoly and Smolkin, Michael",
    title = "{Model of Persistent Breaking of Discrete Symmetry}",
    eprint = "2106.09723",
    archivePrefix = "arXiv",
    primaryClass = "hep-th",
    doi = "10.1103/PhysRevLett.128.011601",
    journal = "Phys. Rev. Lett.",
    volume = "128",
    number = "1",
    pages = "011601",
    year = "2022"
}

@article{Eichhorn:2013mcl,
  title = {Multicritical behavior in models with two competing order parameters},
  author = {Eichhorn, Astrid and Mesterh\'azy, David and Scherer, Michael M.},
  journal = {Phys. Rev. E},
  volume = {88},
  issue = {4},
  pages = {042141},
  numpages = {14},
  year = {2013},
  month = {Oct},
  publisher = {American Physical Society},
  doi = {10.1103/PhysRevE.88.042141},
  url = {https://link.aps.org/doi/10.1103/PhysRevE.88.042141}
}

@article{Chai:2020symmetry,
  title={Symmetry breaking at all temperatures},
  author={Chai, Noam and Chaudhuri, Soumyadeep and Choi, Changha and Komargodski, Zohar and Rabinovici, Eliezer and Smolkin, Michael},
  journal={Phys. Rev. Lett.},
  volume={125},
  number={13},
  pages={131603},
  year={2020},
  publisher={APS},
  doi={10.1103/PhysRevLett.125.131603}
}

@article{Fisher:1974,
  title = {Renormalization-Group Analysis of Bicritical and Tetracritical Points},
  author = {Nelson, David R. and Kosterlitz, J. M. and Fisher, Michael E.},
  journal = {Phys. Rev. Lett.},
  volume = {33},
  issue = {14},
  pages = {813--817},
  numpages = {0},
  year = {1974},
  month = {Sep},
  publisher = {American Physical Society},
  doi = {10.1103/PhysRevLett.33.813},
  url = {https://link.aps.org/doi/10.1103/PhysRevLett.33.813}
}

@article{Vicari:2003,
  title = {Multicritical phenomena in $\mathrm{O}{(n}_{1})\ensuremath{\bigoplus}\mathrm{O}{(n}_{2})$-symmetric theories},
  author = {Calabrese, Pasquale and Pelissetto, Andrea and Vicari, Ettore},
  journal = {Phys. Rev. B},
  volume = {67},
  issue = {5},
  pages = {054505},
  numpages = {12},
  year = {2003},
  month = {Feb},
  publisher = {American Physical Society},
  doi = {10.1103/PhysRevB.67.054505},
  url = {https://link.aps.org/doi/10.1103/PhysRevB.67.054505}
}

@article{Chai:2021tpt,
    author = "Chai, Noam and Dymarsky, Anatoly and Goykhman, Mikhail and Sinha, Ritam and Smolkin, Michael",
    title = "{A model of persistent breaking of continuous symmetry}",
    eprint = "2111.02474",
    archivePrefix = "arXiv",
    primaryClass = "hep-th",
    doi = "10.21468/SciPostPhys.12.6.181",
    journal = "SciPost Phys.",
    volume = "12",
    number = "6",
    pages = "181",
    year = "2022"
}

@article{liu1973quantum,
  title={Quantum lattice gas and the existence of a supersolid},
  author={Liu, Kao-Shien and Fisher, Michael E},
  journal={Journal of Low Temperature Physics},
  volume={10},
  pages={655--683},
  year={1973},
  publisher={Springer},
  doi={10.1007/BF00655458}
}

@article{Hawashin:2025,
  title = {Ultraviolet-Complete Local Field Theory of Persistent Symmetry Breaking in $2+1$ Dimensions},
  author = {Hawashin, Bilal and Rong, Junchen and Scherer, Michael M.},
  journal = {Phys. Rev. Lett.},
  volume = {134},
  issue = {4},
  pages = {041602},
  numpages = {6},
  year = {2025},
  month = {Jan},
  publisher = {American Physical Society},
  doi = {10.1103/PhysRevLett.134.041602},
  url = {https://link.aps.org/doi/10.1103/PhysRevLett.134.041602}
}

@article{KomargodskiPopov:2024,
  title={Temperature-Resistant Order in 2+1 Dimensions},
  author={Komargodski, Zohar and Popov, Fedor K},
  journal={Phys. Rev. Lett.},
  year={2025},
  volume={135},
  number={9},
  pages={091602},
  doi={10.1103/yb7d-6tvc},
  eprint={2412.09459},
  archiveprefix={arXiv}
}

@book{herbut2007modern,
  title={A modern approach to critical phenomena},
  author={Herbut, Igor},
  year={2007},
  publisher={Cambridge University Press},
  doi={10.1017/CBO9780511755521}
}

@article{entropicorder,
	author = {Han, Yiqiu and Huang, Xiaoyang and Komargodski, Zohar and Lucas, Andrew and Popov, Fedor K.},
	date = {2025/11/29},
	doi = {10.1038/s41467-025-66797-3},
	id = {Han2025},
	isbn = {2041-1723},
	journal = {Nature Communications},
	number = {1},
	pages = {87},
	title = {Entropic order},
	url = {https://doi.org/10.1038/s41467-025-66797-3},
	volume = {17},
	year = {2025}}

@article{Hawashin:2025ikp,
  author = {Hawashin, Bilal and Scherer, Michael M. and Smolkin, Michael and Yung, Lev},
  title = {{Spontaneous spacetime parity breaking without thermal restoration}},
  eprint = {2507.19890},
  archiveprefix = {arXiv},
  primaryclass = {hep-th},
  year = {2026},
  journal = {Phys. Rev. D},
  volume = {114},
  number = {6},
  pages = {065005},
  doi = {10.1103/cc6w-syd1}
}

@article{Chai:2020hnu,
  author = {Chai, Noam and Rabinovici, Eliezer and Sinha, Ritam and Smolkin, Michael},
  title = {{The bi-conical vector model at $1/N$}},
  eprint = {2011.06003},
  archiveprefix = {arXiv},
  primaryclass = {hep-th},
  doi = {10.1007/JHEP05(2021)192},
  journal = {JHEP},
  volume = {05},
  pages = {192},
  year = {2021},
}

\end{document}